# Graphene, universality of the quantum Hall effect and redefinition of the SI

T.J.B.M. Janssen[1], N.E. Fletcher[2], R. Goebel[2], J.M. Williams[1], A. Tzalenchuk[1], R. Yakimova[3], S. Kubatkin[4], S. Lara-Avila[4], and V.I. Falko[5]

[1] National Physical Laboratory, Teddington, TW11 0LW, UK
[2] Bureau International des Poids et Mesures, Pavilon de Breteuil, F-92312, France
[3] Department of Physics, Chemistry and Biology, Linköping University, S-581 83 Linköping, Sweden
[4] Department of Microtechnology and Nanoscience, Chalmers University of Technology, S-412 96 Göteborg, Sweden
[5] Physics Department, Lancaster University, Lancaster LA1 4YB, UK

**The Système Internationale d'unités (SI system) is about to undergo its biggest change in half a century by redefining the units for mass and current in terms of the fundamental constants $h$ and $e$, respectively. This change crucially relies on the exactness of the relationships which link these constants to measurable quantities. Here we report the first direct comparison of the integer quantum Hall effect in epitaxial graphene with that in GaAs/AlGaAs heterostructures. We find no difference of the quantized resistance value within the relative standard uncertainty of our measurement of $8.6 \times 10^{-11}$, being the most stringent test of the universality of the quantum Hall effect in terms of material independence.**

The new quantum SI units for mass and current will be based on the fundamental constants of nature $h$, Planck's constant, and $e$, the electron charge. The confidence in the new definition crucially relies on the ability to experimentally confirm the exactness of the relationships which link these constants to measurable quantities. The quantum Hall effect (QHE) defines one such a relationship through the theoretical argument that the Hall resistance is quantized in units of $h/Ne^2$ where $N$ is an integer. The QHE is a fascinating macroscopic quantum effect occurring in two-dimensional conductors that has become one of the cornerstones of the worldwide reference system for scientific and industrial measurements.[1] Yet, the hypothesis of resistance quantization units of $h/Ne^2$ and its independence of material implementation has to be tested experimentally. The appearance of an unusual half-integer variation of the QHE in graphene [2,3] confirmed the unique electrical properties of this 2-dimensional carbon material, where the charge carriers behave as massless Dirac fermions. As well as providing an experimental system for the study of new transport physics, graphene holds out the prospect of a more robust implementation of the QHE resistance standard.[4]

We report here the result of a highest precision direct comparison of the quantized resistance, $R=h/2e^2$, realised in an epitaxial graphene QHE sample with the matching $N=2$ plateau of the QHE in a traditional GaAs/AlGaAs heterostructure device. Demonstrating the equivalence of this resistance in different devices is a vital step in proving the suitability of graphene for metrological use, but is also a useful test of the theory that predicts no corrections to the simple relation $R=h/Ne^2$. The quantum Hall resistance is considered to be a topological invariant, not altered by the electron-electron interaction, spin-orbit coupling, or hyperfine interaction with nuclei, and insensitive to much more subtle influences of gravity.[5] Recently, a quantum electrodymical correction to the von Klitzing constant of the order of $10^{-20}$ has been predicted for practical magnetic field values.[6] The fundamental nature of the Hall resistance quantization makes experimental tests of its universality of the utmost importance, in particular, for improving our knowledge of two fundamental quantities of nature: the electron charge and Planck's constant. The precision obtained through a universality test as presented here is much greater than is possible by a comparison to the values of the constants $h$ and $e$.[7] Analysis of the complete set of published results carried out by CODATA [7] showed no deviation from $h/e^2$ to within $2\times10^{-8}$, which calls for more accurate measurements.

Soon after the first observations of the QHE in graphene[2,3], Giesbers et al.[8] reported an evaluation of the accuracy of the resistance quantization in exfoliated graphene flakes. Unfortunately, the small size of the flakes and electrical contacts along with the low breakdown current in their devices made these measurements very difficult. An accuracy of only a few parts in a million could be obtained (4 orders of magnitude below the state-of-the-art in GaAs and Si) and so no meaningful conclusions on the universality of the QHE could be drawn. Our own work in Ref. [9] reported the first accurate observation of the QHE in large epitaxial graphene devices. We achieved an accuracy of 3 parts in $10^9$ via an indirect method whereby both quantum Hall devices were separately measured against a room temperature standard resistor. Recently, we reported[10] an unusually strong pinning of the $\nu=2$ quantum Hall state in epitaxial graphene due to charge exchange with the localised states in the substrate resulting in a very robust resistance quantization and demonstrated invariance of the resistance quantization to 0.3 parts in $10^9$ over a field range of 3.5 T. Importantly for precision metrology, the extraordinarily robust quantum Hall state in these devices sustains very high non-dissipative currents ensuring a large signal-to-noise ratio.

Our graphene sample was produced by epitaxial growth on a SiC substrate [9] and shows the properties (such as low contact resistance and negligible longitudinal resistivity) required for accurate metrological use. Its resistance was compared to that of the GaAs device in a null measurement using the standard methods of resistance metrology. (The 4-terminal nature of QHE resistors means that some form of bridge circuit is needed, even to compare identical resistors; here a cryogenic current comparator [11] was used to establish an exact 1:1 current ratio.) A summary of the results is shown in Figure 1 (for details see Methods section).

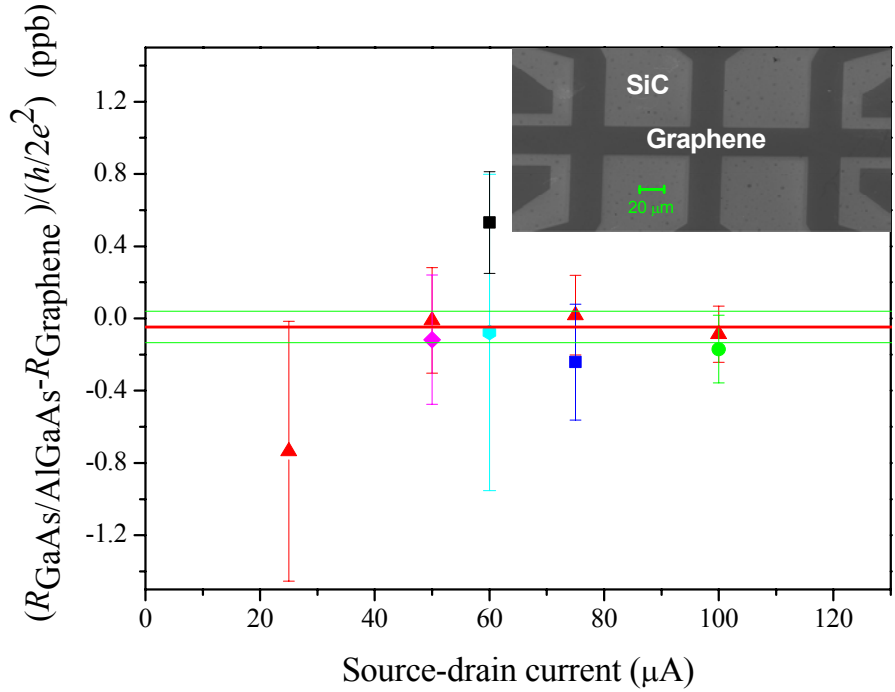

*Figure 1 shows the measured difference between a GaAs/AlGaAs sample and a graphene sample as function of the source-drain current through the devices for different measurement configurations. Red triangles: GaAs/AlGaAs device 1 in system 1 at 1.5K – graphene in system 2 at 300mK, Green dot: measured using non-opposite voltage contacts on GaAs/AlGaAs device 1. Blue square: measured using non-opposite contacts on graphene device. Pink diamond: GaAs/AlGaAs device 2 – graphene. Light blue hexagon is measured in reverse magnetic field for graphene. Black square: samples exchanged between system 1 and 2. The red line is the weighted mean of all the data points and the green lines signify +/-1 standard deviation. Inset: SEM picture of graphene device.*

The weighted average of all our data is $(R_{GaAs/AlGaAs}-R_{Graphene}) / (h/2e^2) = (-4.7 \pm 8.6) \times 10^{-11}$. The relative standard uncertainty of $8.6 \times 10^{-11}$ represents a factor of 35 improvement on our prior result obtained via an indirect measurement.[9,10] In an indirect measurement the accuracy is limited by the properties of the resistor used as a transfer standard.[1] Here we directly compare both devices against each other thereby

---

[1] Note an important distinction between the precision of the measurement and the accuracy of the result. Precision is used to define the measurement repeatability, whereas accuracy expresses how close

eliminating many systematic effects. Previously our knowledge of the universality of the QHE has been limited to the level of 2 or $3\times10^{-10}$ for comparisons between GaAs and Si or between identical GaAs devices.[1] However both GaAs and Si are traditional semiconductors with a parabolic bandstructure and governed by the same physics. Graphene is a semimetal with a linear bandstructure and is described by Dirac-type massless charge carriers and so universality in terms of material independence goes well beyond the comparison between two semiconductors. In our universality experiment the maximum source-drain current that the GaAs device can sustain without dissipation limits the measurement uncertainty, whereas a potentially lower uncertainty can be obtained in a consistency check of two graphene devices.

Our results on material independence is the strongest evidence yet of the hypothesis that the resistance is quantized in units of $h/Ne^2$ is correct and thereby supports the pending redefinition of the SI-units for kilogram and ampere in terms of $h$ and $e$.[12] Judging from the robustness of the quantization and wide operational parameter space, epitaxial graphene should be the material of choice for quantum resistance metrology.

**Methods**
The epitaxial graphene sample used in the reported experiment was produced on the Si-face of SiC.[9] The graphene Hall bar was encapsulated in a polymer bilayer, a spacer polymer followed by an active polymer able to generate acceptor levels under UV light. More fabrication details can be found elsewhere.[13] The sample had an electron density, $n_S$, of $4.6\times10^{11}$ cm$^{-2}$ and mobility, $\mu$, of 7500 cm$^2$V$^{-1}$s$^{-1}$. Note that this mobility is rather low compared to that achieved in exfoliated or suspended graphene and much lower than that obtained in the best GaAs. Fortuitously, in the quantum Hall effect disorder is in fact necessary to provide localisation of the electron states and for precision metrology the mobility should not be too high in order to provide a wide quantum Hall plateau. A standard 8-contact Hall bar geometry was patterned on the device with dimensions 160 μm x 35 μm. The graphene sample was placed in system 1 at 300 mK and 14 Tesla. The two GaAs samples used were traditional GaAs-AlGaAs heterostructures obtained from the PTB (device 1) and LEP (device 2). Device 1 had $n_S = 4.6\times10^{11}$ cm$^{-2}$ and $\mu = 4\times10^5$ cm$^2$V$^{-1}$s$^{-1}$, the size of the chip was 6000 μm x 2500 μm and contacts were made from small tin balls at the edge of the chip. Device 2 had $n_S=5.1\times10^{11}$ cm$^{-2}$, $\mu = 5\times10^5$ cm$^2$V$^{-1}$s$^{-1}$; the chip had an etched Hall-bar geometry of 2200 μm x 400 μm and AuNiGe alloyed contacts. Both GaAs devices were placed in system 2 at 1.5 K and either 9.5 T (device 1) or 10.5 T (device 2). Before commencing the high-accuracy measurements all devices were fully characterised according to the guidelines for quantum Hall resistance metrology [1] (i.e. we confirmed that the three-terminal contact resistance measured on the $N=2$ plateau was of the order of a few ohms for all contacts used and that the longitudinal resistivity at the measurement current was below 10 μΩ). For the graphene device the maximum source-drain current, $I_C$, at which the device remains in the non-dissipative state was approximately 500 μA. For the GaAs devices $I_C$ was ≈150 μA for device 1 and ≈100 μA for device 2.

---

is the measured value from the true value. [International Vocabulary of Metrology, http://www.bipm.org/en/publications/guides/vim.html]

The measurements were made with a cryogenic current comparator (CCC) bridge [11] illustrated in simplified form in Fig. 2. Isolated current sources 1 and 2 separately drive current through samples S1 and S2 and associated windings A and B on the CCC. The current ratio can be set via electronics to a few parts in $10^6$ and this ratio is improved to a level of 1 part in $10^{11}$ by forming a negative feedback loop from the SQUID sensing the net flux in the CCC to one of the current sources. The potential contacts on S1 and S2 are closed in a loop via winding C on a second CCC. This device is configured with just a single winding to measure a current null rather than two windings to establish a current ratio. Data are collected alternately in forward and reverse current direction so as to eliminate electrical offsets. Measurement uncertainty arises from leakage currents in the connecting cables, residual error in the ratio A:B, accuracy of the negative feedback loop and random noise. The random noise of 8.7 parts in $10^{11}$ dominates over the other components, estimated to have a combined standard uncertainty of 1.6 parts in $10^{11}$.

**Acknowledgments**
The authors thank Stephen Giblin and Dale Henderson for their assistance. The work was supported by the UK National Measurement Office (NMO) Pathfinder Programme, Swedish Research Council and Foundation for
Strategic Research, EU FP7 STREPs ConceptGraphene and SINGLE, EPSRC grant EP/G041954 and the Science \& Innovation Award EP/G014787.

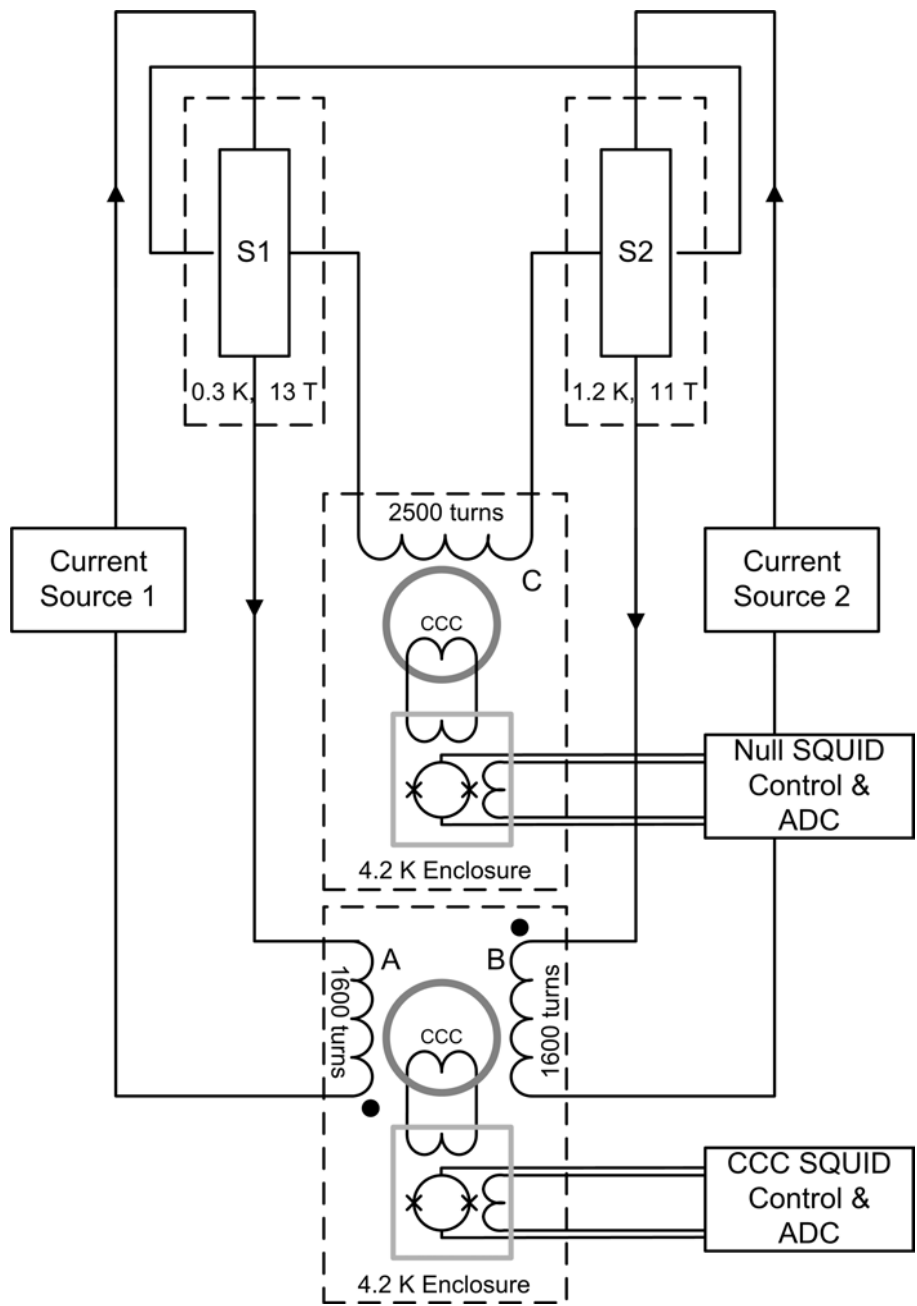

**Figure 2** *Simplified schematic of the cryogenic current comparator bridge circuit.*